\newcounter{myctr}

\documentclass{ws-ijqi}

\begin{document}

\markboth{H.-T. Elze}{Cellular automaton ontology, bits, qubits, and the Dirac equation}


\title{CELLULAR AUTOMATON ONTOLOGY, BITS, QUBITS, AND THE DIRAC EQUATION}

\author{Hans-Thomas Elze}

\address{Dipartimento di Fisica ``Enrico Fermi'', Universit\`a di Pisa, \\ 
Largo Pontecorvo 3, I-56127 Pisa, Italia \\ 
elze@df.unipi.it}

\maketitle


\begin{abstract}
Cornerstones of the {\it Cellular Automaton Interpretation of Quantum Mechanics} are its ontological states that evolve by permutations, in this way never creating would-be quantum mechanical superposition states. We review and illustrate this with a classical Ising spin chain. It is shown that it can be related to the Weyl equation in the continuum limit. Yet, the model of discrete spins or bits unavoidably becomes a model of qubits by generating superpositions, if only slightly deformed. We study modifications of its signal velocity which, however, do not relate to mass terms. To incorporate the latter, we consider the Dirac equation in 1+1 dimensions and sketch an underlying discrete deterministic ``necklace of necklaces'' automaton that qualifies as ontological. 
\end{abstract}



\section{Introduction -- The Big Picture}  

Let us outline a still hypothetical deterministic ontology that may be associated with the natural phenomena which are studied in physics and chemistry, if not beyond. This is, of course, in contrast with the perception that there is an intrinsic fundamental randomness seen, in particular, in observations of or experimentation with the dynamics of microscopic objects, such as Standard Model `elementary particles', atoms, or molecules, see, {\it e.g.}, the  references.~\cite{GisinBook,Haag2013,Haag2016} This widespread idea is obviously founded on a century of extremely successful developments and applications of quantum theory in the description and manipulation of the microscopic world which surrounds and affects us.   

Unlike recent studies of {\it quantum cellular automata} or {\it quantum walks}, which may be of interest for the foundations of (quantum) physics or in (information processing) applications, presently we do {\it not} base our discussion on quantum theory from the outset. Although we will quickly find it convenient to borrow elements from its very efficient formalized language. 
This resonates with our working hypothesis that what we observe in experiments, we `see' in terms of descriptions employing the language of mathematics and especially of quantum theory. 

Following the 
{\it Cellular Automaton Interpretation (CAI) of Quantum Mechanics} proposed by 't\,Hooft,~\cite{tHooft2014} 
we shall consider circumstances when quantum mechanical features are found in the behaviour of certain kinds of classical cellular automata (see also a recent article \cite{tHooft2023} for related references). This posits deterministic dynamics to be responsible for any change occurring with an object, corresponding to a change of the ontological state.   

The Universe is always in a specific, even if unknown to us, ontological state. In the `toy models' considered in the following, the model universe consists in a countable number of two-state Ising spins that may form a one-dimensional chain and evolve according to deterministic automaton rules. -- Incidentally, the yes--no or spin-up--spin-down alternatives incorporated here can be described as occupation numbers of fermionic degrees of freedom, {\it i.e.} with values 0 or 1. The latter has been extensively explored by Wetterich developing fermionic quantum field theories from classical {\it probabilistic cellular automata}.\,\cite{WetterichDICE22,Wetterich22} 

Presently, we shall stick to strictly deterministic evolutions of cellular automata, as in earlier work.\,\cite{PRA2014,Wigner13} 
However, we will base this on and elaborate more recent results, which include a wider discussion of the motivation to reexamine the foundations of quantum theory in the light of classical concepts -- especially determinism and existence of ontological states of reality -- with numerous related references.\,\cite{Elze22,Elze20,Elze20b}

Thus, we will not enter anew debates surrounding attempts to reinstall determinism at a fundamental level of natural science. Likewise with related arguments concerning the measurement problem of quantum theory, which in an ontological theory does not exist,\,\cite{tHooft2014} and the violation of Bell's inequalities, which not necessarily rules out such a theory.\,\cite{tHooft2014,Vervoort1,Vervoort2,Wetterich20,Hossenfelder,Vervoort3} 

Instead we discuss interesting features of ontological states realized in classical Ising models. -- We may picture such states simply as a collection of {\it numbered either black or white pebbles} sitting in fixed single-pebble holes. Their deterministic dynamics consists in unitary permutations which amounts to reshuffling of pebbles, {\it e.g.} as in the following transition of an in- to an out-state: 
$$|in\rangle \equiv\;\;  \stackrel{1}{\bullet}\stackrel{2}{\bullet}\stackrel{4}{\bullet}\stackrel{3}{o}\stackrel{5}{o}\stackrel{6}{\bullet}\;,
\;\; |out\rangle \equiv\;\;  \stackrel{2}{\bullet}\stackrel{6}{\bullet}\stackrel{5}{o}\stackrel{1}{\bullet}\stackrel{3}{o}\stackrel{4}{\bullet} 
\;\; =: \hat U|in\rangle\;,
\;\; \hat U\hat U^\dagger =\hat U^\dagger\hat U=\mathbf{1}\;.$$ 	
Generally, then, our task is to study the Hamiltonian operator $\hat H$ defined by: 
$$\hat U(T)=: e^{-i\hat HT}\;,$$ 
where $T$ is the elementary unit of time that it takes for one update of the state, as illustrated in the above example. Incidentally $T$ sets also the energy scale for the Hamiltonian (choosing units such that $\hbar =c=1$ henceforth).  

Several remarks are in order here. (Note that we are conveniently using mathematical terminology that is  familiar from quantum theory.) -- First of all, there are {\it no superposition states} created by any dynamics generated through permutations.\,\footnote{We will recall in the next section that absence of reduction or `collapse' of the wave function (stipulated by the measurement problem) is ultimately due to absence of superposition states.} 

While superpositions do not belong to a set of ontological states, it has turned out to be convenient to introduce them in the mathematical language used in doing physics, the {\it quantum superpositions} of ontological (micro) states, which do not exist `out there' as states the Universe can possibly be in. In distinction, {\it  classical states} are ontological (macro) states, or probabilistic superpositions thereof, appropriate for nature's phenomena at vastly different scales.~\cite{tHooft2014}  

The approach referred to here does not aim to replace quantum theory as the physicist's perfect toolbox. However, one would like to better understand 
its pecularities. It seems desirable to recover the structural elements of its mathematical language through coarse-graining and a suitable continuum limit of an ontological picture.~\cite{PRA2014} See also  Refs.\,\cite{tHooft2014,IJQI17,FFM17} for further discussions, as well as other more and less related attempts.~\cite{H1,H2,H3,Elze,Kleinert,Groessing,Margolus,Jizba,Mairi,Isidro,DArianoCA,Wetterich16a} 
We point out in particular also three rather evolved, though not uncontested approaches to quantum theory,  
namely Bohmian mechanics\,\cite{Bohm52,Tumulka18,Gisin21}, stochastic electrodynamics\,\cite{Santos22,Boyer19,CettoEtAl15}, and prequantum classical statistical field theory\,\cite{Khrennikov1,Khrennikov2}, 
studying its emergence or generalization under various perspectives. 
 
Before introducing a more detailed `toy model', we draw already attention to the fact that models based on permutations of ontological states lead to Hamiltonians which are rather poor in tunable parameters or coupling constants. Presently, besides the chain or other discrete geometry, its dimensionality, connectivity etc., only the scale $T$ is available. In the long run of developing a future ontological theory this may prove to be a desirable feature.  

For our present purposes it suffices to point out that a  Hamiltonian pertaining to the unitary dynamics generated by permutations will be completely fixed, once an underlying Ising model with its automaton updating rule and time step $T$ is defined. However, interpreting this with an eye on realistic circumstances of a physicist, who tries to pin down a theoretical model based on experimental data, we notice that inaccuracies of extracted parameters cannot be avoided. We have argued that such unavoidable imprecision will replace the searched for Hamiltonian $\hat H$ -- describing  deterministic evolution of ontological states -- by a result based on experiments, $\hat H_{exp}=\hat H+\delta\hat H$. And that most likely $\hat H_{exp}$ will generate superpositions and, thus, open a Hilbert space, the arena of quantum theory.\,\cite{Elze20}

In Section\,2., we briefly recall the distinction between ontological, classical, and quantum states that we refer to throughout this paper. 

In Section\,3., we present our study of a particular 
Ising model with two-state `classical spins' which evolve deterministically and linearly by permutations. We shall see that the corresponding update dynamics proceeds with a final signal velocity on or within discrete light-cones. 
This may be considered as a typical `toy model' for the cellular automaton ontology. Namely, we argue that `by mistake' this discrete model becomes a model of interacting qubits, {\it i.e.} quantum mechanical behaviour is seen as the outcome of approximations or an inaccurate knowledge of the underlying 
ontological states and their dynamics. We obtain equations of motion of left- and right-movers and show that they are related to the Weyl equation for massless particles in the continuum limit. 

In order to introduce mass, we consider in Section\,4. the Dirac equation in 
1+1 dimensions and outline a deterministic automaton model based on its discretized version. Which figuratively may be called a {\it ``Necklace of Necklaces''}. Also this model qualifies as ontological, since its discrete states evolve by permutations without creating superpositions and, thus, without 
invoking quantum mechanics. -- Conclusions follow in Section\,5. 


\section{Basic ingredients of deterministic automaton models}

For completeness, we recall here some elementary notions 
in terms of which ontological cellular automaton models can be formulated. 

To begin with, we emphasize once more that quantum states will be considered to form part of the mathematical language used and, thus, bear an epistemological character.~\cite{Rovelli2015} They are ``templates'' for describing physical reality, which consists of ontological states ruled by deterministic  evolution.~\cite{tHooft2014}  

{\it Ontological States} ($\cal OS$) are the physical states a closed physical system can be in. For simplicity, we assume that the set of this states is denumerable. -- 
{\it No Superpositions} of $\cal OS$ exist `out there'. 

{\it Permutations} evolve $\cal OS$ into $\cal OS$. Denoting them by $|A\rangle ,\; |B\rangle ,\; |C\rangle ,\; |D\rangle ,\;\dots\;$, the evolution could be as follows, for example:   
$$|A\rangle\rightarrow |C\rangle\rightarrow |B\rangle\rightarrow|D\rangle\rightarrow\;\dots\;\;.$$  
Note that this kind of evolution is the only possible one, unless the set of states itself changes, {\it i.e.}, grows or shrinks. 

{\it Hilbert space} ${\cal H}$ can be founded 
on the $\cal OS$, once they are declared to form an  orthonormal set. This distinct primordial basis is fixed once for all. -- Diagonal operators on this basis are {\it beables} and their eigenvalues characterize physical properties of the states, corresponding to the labels $A,\; B,\; C,\;\dots\;$ above. 

{\it Quantum States} ($\cal QS$) are then introduced as superpositions of $\cal OS$ in ${\cal H}$. -- They belong to the mathematical language of quantum theory. -- Consider, {\it e.g.}:  
$$|Q\rangle :=\alpha |A\rangle +\beta |B\rangle +\dots\;,
\;\;|\alpha |^2+|\beta |^2 +\dots\; =1\;.$$
The complex amplitudes that enter here need to be interpreted, when analyzing experiments with the help of such states. By experience, relating probabilities to these amplitudes has been an extraordinarily useful invention. While the {\it Born rule} is here built in by definition, this relation could well be chosen differently, at the cost of complicating the mathematical tools, as has been discussed before.~\cite{tHooft2014,FFM17} 

An important consequence of the presented set-up is a 
conservation law, the {\it Conservation of Ontology}. 
Staying with the above examples, the state $|Q\rangle$ 
evolves by the permutations like this: 
$$|Q\rangle\rightarrow\alpha|C\rangle +\beta|D\rangle + \dots\;,$$ 
where the amplitudes are constant, retaining their initial values. This means, they reflect probabilities introduced to describe the initial state. As we do not know the initial ontological state for any kind of experiment (embedded in the Universe), we resort to the probabilistic approximate approach perfected in quantum theory.  

We shall see in the next section that -- besides perhaps an only approximate knowledge of the initial state -- an approximate knowledge of the Hamiltonian forces the quantum mechanical description of the evolving object upon us. 

{\it Classical States} ($\cal CS$) are defined in relation to $\cal OS$. -- They have commonly been thought to describe certain limiting situations of QM. However, by {\it CAI} they are associated with deterministic macroscopic systems, such as billiard balls, pointers of measurement devices, planets, and so on.~\cite{tHooft2014} They represent $\cal OS$ that are not resolved individually, but rather lumped together in probabilistic distributions. 

If an experiment is repeatedly performed or an object repeatedly undergoes the same evolution, then, being part of the Universe and not perfectly isolated from the rest or closed, each time different initial conditions regarding the $\cal OS$ are realized. 
A suitable {\it classical apparatus} forming part of such situations must therefore be expected to yield different pointer positions as outcomes each time, with a frequency that reflects the implicitly assumed probability of initial $\cal OS$, entering the description in terms of $\cal QS$.  

Thus, reduction or collapse to a $\delta$-peaked distribution of pointer positions, the core of the {\it measurement problem}, is an apparent effect. It is caused by an intermediary use of $\cal QS$, in particular superposition states,  to describe the evolution of what in reality are $\cal OS$ that differ in different runs of an experiment. 

We learn that an ontological theory may show that QM {\it per se} does not need any stochastic or nonlinear reduction process, which is supposed to modify the collapse-free linear unitary evolution. According to {\it CAI}, measurements are simply {\it interactions} between the degrees of freedom belonging to an object and those belonging to an apparatus, altogether (with the rest of the Universe) evolving through ontological states. 

Our task in the following is to further study some idealized  many-body systems in line with these quite simple concepts, 
with an eye towards more realistic scenarios. 


\section{Left- and right-moving Ising spins} 
Before turning to an Ising spin model, we first review briefly the {\it Cogwheel Model}, which has found a number of applications in contexts related to ontological model building. We mention that this model can describe the quantum mechanics of a free particle or an harmonic oscillator in suitable (continuum) limits.\,\cite{tHooft2014,tHooft2023,Elze,ElzeRelativPart} 


\subsection{The Cogwheel Model} 

The Cogwheel Model is perfectly suited to illustrate the 
dynamics generated by permutations of some simple `ontological states'.\,\footnote{Here we follow Refs.\,\cite{Elze22,Elze20}, where more details can be found.} 
 
As {\it complete permutations} of a set of $N$ objects, say $A_1,A_2,\dots ,A_N$, the $\cal OS$, we consider a bijective mapping of {\it all} states in $N$ steps onto one another. 
This can always be represented by an {\it unitary} $N\times N$ 
matrix $\hat U$ with one off-diagonal phase per column and 
row and zeros as entries elsewhere: 
\begin{equation}\label{permutation} 
\hat U_N:= 
\left (
\begin{array}{c c c c c} 
0           & .           & .     & 0       & e^{i\phi_N} \\ e^{i\phi_1} &0            &       & .       & 0  \\ 
0           & e^{i\phi_2} & .     &         & .  \\
.           &             & .     & .       & .  \\
0           & .           & 0     & \;e^{i\phi_{N-1}} & 0  \\
\end{array}\right )
\;\;. \end{equation}
This refers to what we call {\it standard basis}, {\it i.e.} having suitably ordered the states. 

Since $(\hat U_N)^N=e^{i\sum_{k=1}^N \phi_k}\;\mathbf{1}\;$, 
one immediately finds the eigenvalues of this matrix as the 
$N$\,th roots of 1, multiplied by an overall phase, which lie on a unit circle in the complex plane. -- We set the phases $\phi_k$ to zero, since they will play no role in what follows. 

We now define the Hamiltonian operator $\hat H$ that corresponds to $\hat U$ by: 
\begin{equation}\label{Hamiltonian} 
\hat U_N=:e^{-i\hat H_NT} 
\;\;, \end{equation} 
where $T$ is the time scale, respectively $T^{-1}$ the energy scale, mentioned in the Introduction. The eigenvalues of $\hat H$ and its matrix elements are then easily found:  
\begin{equation}\label{Hamiltoniandiag} 
(\hat H_N)_{nn}=\frac{2\pi}{NT}(n-1)
\;,\;\;n=1,...,N
\;\;, \end{equation}
with respect to what we call the {\it diagonal basis}. 

Since the standard and diagonal bases are related by a unitary discrete Fourier transformation, the matrix elements of the Hamiltonian with respect to the diagonal basis can also be calculated. This yields: 
\begin{eqnarray}\label{Hstandardond} 
(\hat H_N)_{nn}&=&\frac{\pi}{NT}(N-1)\;,\;\; n=1,\dots ,N\;\;, 
\\[1ex]\label{Hstandardoffd}
(\hat H_N)_{n\neq m}&=&\frac{\pi}{NT}
\Big (-1+i\cot\big (\frac{\pi}{N}(n-m)\big ) \Big ) 
\;,\;\; n,m=1,\dots ,N  
\;\;, \end{eqnarray} 
and $\hat H_N=\hat H_N^\dagger$, as it should. 


\subsection{The Ising spin chain with pairwise exchange interactions} 

Our model consists of $2S+1$ `classical' two-state Ising spins -- corresponding to Boolean variables or fermionic occupation numbers -- forming a one-dimensional chain with periodic boundary condition ({\it p.b.c.}).\,\cite{Elze22,Elze20} Thus, the chain is always in one of 
$2^{2S}$ states, the $\cal OS$:
\begin{equation}\label{chainstate} 
|\psi\rangle :=|s_1,s_2,\dots,s_{2S};s_{2S+1}\equiv s_1\rangle 
\;\;, \end{equation}
where the spin variables are labeled according to the ordered positions along the chain and can assume the values 
\begin{equation}
\label{Isingspins} 
s_k=\pm 1\;,\;\; k=1,\dots ,2S+1\;,\;\;s_{2S+1}\equiv s_1\;\;(p.b.c.) 
\;\;. \end{equation} 
The permutations of the $\cal OS$ are generated by 		
transpositions, {\it i.e.} exchange of a pair of spins, with the following properties:  
\begin{equation}\label{spinexchange} 
\hat P_{ij}|s_i,s_j\rangle :=|s_j,s_i\rangle\;, 
\;\;\hat P_{2S\;2S+1}\equiv\hat P_{2S\;1}
\;\;, \end{equation}
and, 
\begin{equation}\label{exchangeprop} 
\hat P^2=\mathbf{1}\;,
\;\;[\hat P_{ij},\hat P_{jk}]\neq 0\;, 
\;\;\hat P_{ij}= (\underline{\hat\sigma}_i\cdot\underline{\hat\sigma}_j+\mathbf{1})/2 
\;\;, \end{equation} 
where the last relation shows the well known representation 
in terms of the vector of Pauli matrices, 
$\underline{\hat\sigma}:=(\sigma_x,\sigma_y,\sigma_z)^t$. 
Note that $\hat P_{ij}=\hat P_{ji}$, with the indices referring to fixed sites   

Then, the evolution of $\cal OS$ is determined by the unitary update operator $\hat U$: 
\begin{equation}\label{Uchain}
\hat U:=\prod_{k=1}^S\hat P_{2k-1\;2k}\prod_{l=1}^{S}\hat P_{2l\;2l+1}
=:\exp (-i\hat HT)
\;\;, \end{equation}
to which is related the chain Hamiltonian $\hat H$, introducing the time step $T$, as discussed before. We grouped the sequence of pairwise exchanges according to whether they concern either {\it even} or {\it odd pairs} $ij$, defined by the first index $i$ of $\hat P_{i<j}$ being either even or odd. 

A number of properties of such Ising spin chains have been discussed earlier.\,\cite{Elze22,Elze20,Elze20b} Here we 
represent the extracted Hamiltonian $\hat H$ corresponding to the update operator $\hat U$ and continue to discuss the dynamics of {\it left-movers} and {\it right-movers} of related decoupled `cogwheels' ({\it cf.} subsection\,3.1.).   

The operator $\hat H$ pertaining to the operator 
$\hat U$, {\it cf.} Eq.\,(\ref{Uchain}), is given by:  
\begin{eqnarray}\label{HamiltonianCog} 
\hat H&=&\sum_{n=1}^S(\hat H_S)_{n1}\hat U^{n-1} 
\\ [1ex] \label{HamiltonianNonStat} 
&=&\frac{\pi}{T}\Big (\mathbf{1}
+\frac{i}{S}\sum_{n=1}^{S-1}\cot (\frac{\pi}{S}n )\hat U^{n}\Big )
\;,\;\;\mbox{for}\;\;\hat U|\psi\rangle\neq |\psi\rangle 
\\ [1ex]\label{HamiltonianFin}   
&=&\frac{\pi}{T}\Big (\mathbf{1}
+\frac{i}{2S}\sum_{n=1}^{S-1}\cot (\frac{\pi}{S}n )
\big (\hat U^{n}-(\hat U^\dagger )^{n}\big )\Big )
\;\;, \end{eqnarray}
where matrix elements of an $S$-state cogwheel Hamiltonian, 
Eqs.\,(\ref{Hstandardond})-(\ref{Hstandardoffd}), appear in 
Eq.\,(\ref{HamiltonianCog}). In Eq.\,(\ref{HamiltonianNonStat}) invariant states are excluded ({\it e.g.} all spins up); for them, of course, one finds $\hat H|\psi_0\rangle =0$. -- Underlying these results is a terminating Baker-Campbell-Hausdorff formula.\,\cite{Elze20b}  

At this point, we stress that the unitary operator $\hat U$ and the hermitean operator $\hat H$ describe nothing but the deterministic evolution of the `classical' Ising spin chain, 
despite their appearance and `quantum language' used. 
{\it No superpositions} of its ontological states are formed ever. We shall come back to this fact. 


\subsection{Left- and right-movers} 

Deriving the Hamiltonian was facilitated by knowing the Cogwheel Model and observing the simple evolution of a generic spin chain state $|\psi\rangle$, Eq.\,(\ref{chainstate}), 
generated by repeatedly applying $\hat U$.\,\cite{Elze22,Elze20}  

Spin variables $s_{2k-1}$ ($s_{2k}$) initially associated with {\it odd}  ({\it even}) {\it numbered sites} 
consecutively move always two steps to the left (right) to the next lower (higher) numbered odd (even) site. These are named  {\it left-movers} ({\it right-movers}). 
  
The periodic boundary condition then leads to periodic 
evolution of the whole chain. Updating the state $S$ times 
leads back to the initial state, 
$\hat U^S|\psi\rangle =|\psi\rangle\;$. Thus, each spin variable performs the motion of a cogwheel, jumping along the even or odd sites of the chain. 

By separating the pairwise exchange interactions of spins as in Eq.\,(\ref{Uchain}) into two groups, with transpositions in each group commuting with each other, a {\it finite signal velocity} has been implemented. The left- and right-movers propagate on some kind of {\it (1+1)-dimensional light-cones}, 
which is illustrated in Figure\,1\,a). 

\begin{figure}
\includegraphics[width=1.0\columnwidth 
]{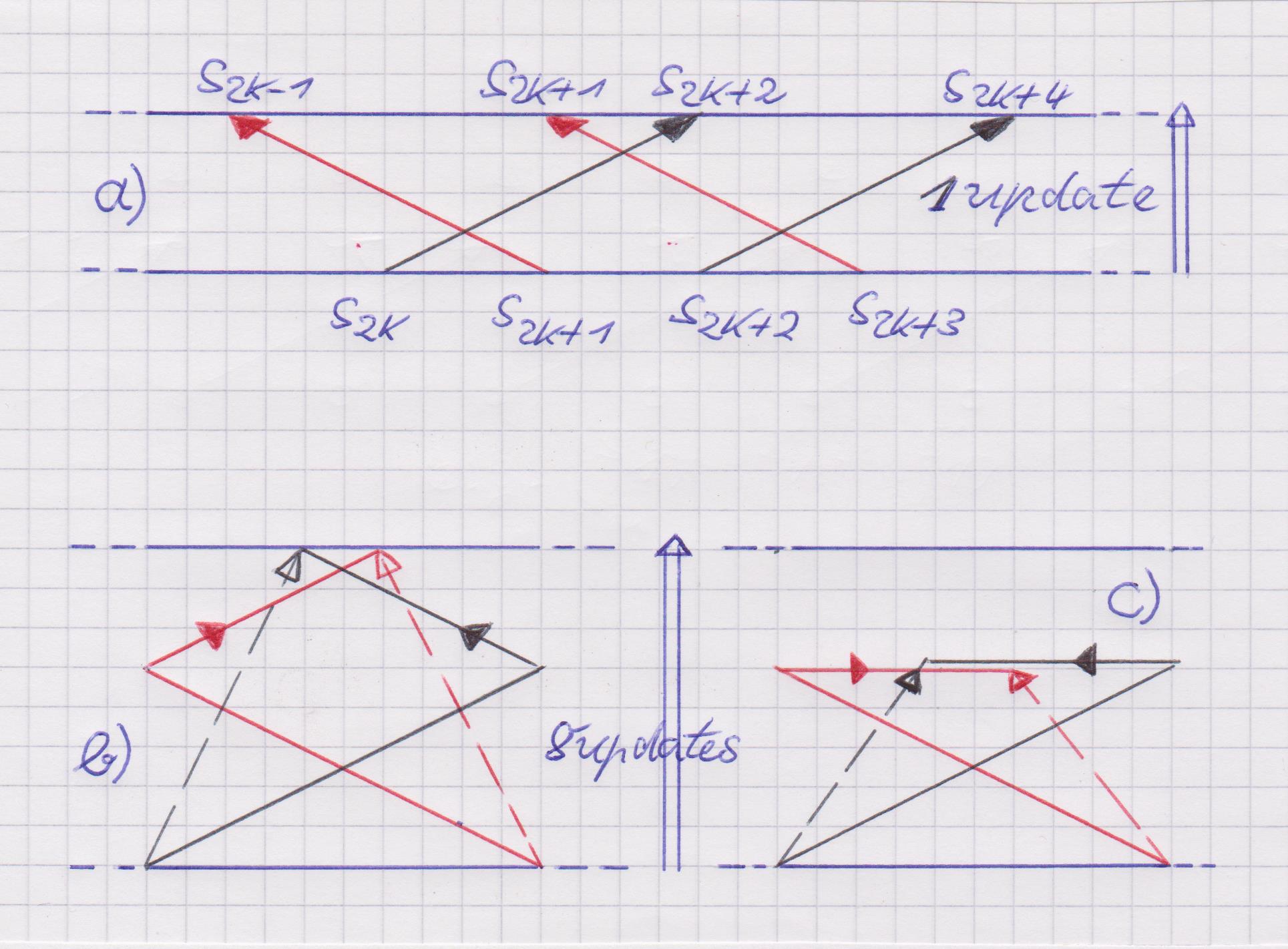}   
\caption{a) Evolution of Ising spins with left- and right-movers, respectively, on odd and even sites of a  chain (periodic boundary conditions; time runs upwards). -- b) \underline{\it Case A} of slowing-down the signal velocity, {\it e.g.} 5 updates followed by 3 with reversed propagation and resulting evolution (dashed lines). -- c) \underline{\it Case B} of slowing-down the signal velocity, {\it e.g.} 5 updates followed by 3 instantaneous translations and resulting evolution (dashed lines). -- See main text for details.} 
\end{figure} 

One may wonder, whether the signal velocity can be modified somehow. Indeed, this can be done, for example, as follows. 

{\it \underline{Case A}}: Suppose we apply the update operator $k_0$ times, $(\hat U)^{k_0}$. This moves a right-mover (left-mover) $2k_0$ sites to the right (left). Since 
$\hat U^\dagger\hat U =\mathbf{1}=\hat U^\dagger \hat U$, we can undo, say, $l_0$ of the moves by applying 
$(\hat U^\dagger )^{l_0}$. Since $\hat U$ and $\hat U^\dagger$ commute, the overall resulting translation is the same, no matter how the sequence of updates and their inverses is distributed among the $k_0+l_0$ steps. An example is shown in Figure\,1\,b).\,\footnote{The phases of reversed motion, which 
slow down the evolution, could also be distributed locally  varying along a chain. This has been considered in  Ref.\,\cite{Wetterich20} to introduce an external scalar potential for a probabilistic automaton, intended to describe  a quantum mechanical particle in a suitable continuum limit.}    
 
For a long chain, $S\gg k_0+l_0$, it might be useful to 
present such slowing-down by an effective  Hamiltonian, $\hat H_{eff}:=\hat H(k_0-l_0)/(k_0+l_0)$, and corresponding update operator, $\hat 
U_{eff}(T_{eff}):=\exp (-i\hat H_{eff}T_{eff})$, with rescaled 
`unit' time step $T_{eff}:=(k_0+l_0)T$. This gives   
$\hat U_{eff}(nT_{eff})=\exp (-in(k_0-l_0)\hat HT)\;$, which shows the slowing-down compared to the case without motion 
reversal, with $n(k_0+l_0)$ in the exponent instead.  

{\it \underline{Case B}}: An alternative would be to have $\hat U^{k_0}$ followed by an 
instantaneous translation by $2l_0$ sites to the left (right)  
for the right-movers (left-movers), with $l_0\leq k_0$. Such a
left translation is obtained by applying $l_0$ times a 
translation by 2 sites to the left, $\hat T_L(2)$, which can 
be generated by transpositions: 
\begin{equation}\label{translationL} 
\hat T_L(2):=\prod_{j=1}^S \hat P_{2j+2\;2j}
\;\; \end{equation} 
with the $\hat P_{j'j}$ ordered from right to left with increasing indices, subject to the periodic boundary condition. Thus, $\hat T_L(2l_0)=(\hat T_L(2))^{l_0}$. 
Analogously, translations to the right are generated by $\hat T_R(2)$: 
\begin{equation}\label{translationR} 
\hat T_R(2):=\prod_{j=1}^S \hat P_{2j-1\;2j+1}
\;\;, \end{equation}  
with ordering from left to right with increasing indices. 
We note that right and left translations commute with each other, since they act on left- and right-movers, respectively, 
while not all transpositions defining one or the other commute among themselves.   

Furthermore, it becomes obvious when describing the actions of 
all three operators on a generic state of the Ising chain 
pictorially - {\it cf.} Figure\,1\,c) - that $\hat U$, $\hat T_L$, and $\hat T_R$ all commute. Therefore, we can find the 
additive contributions of the translations to the Hamiltonian 
$\hat H$ of 
Eqs.\,(\ref{HamiltonianCog})-(\ref{HamiltonianFin}) by representing them as exponentials. This is achieved in the 
same way as extracting $\hat H$ from $\hat U$ before.\,\cite{Elze22,Elze20} Namely, since 
$\hat T_{L,R}(2S)=\mathbf{1}$, by {\it p.b.c.}, also the translations generate $S$-state cogwheel like evolution of right- and left-movers, respectively. Thus, setting 
$\hat T_{L,R}(2)=:\exp (-i\hat\Theta_{L,R}D)\;$ (with distance scale $D$; recall $\hbar =c=1$), we find: 
\begin{eqnarray}  
\Theta_{L,R}
&=&\frac{\pi}{D}\Big (\mathbf{1}
+\frac{i}{2S}\sum_{n=1}^{S-1}\cot (\frac{\pi}{S}n )
\big (\hat T_{L,R}(2)^{n}-(\hat T_{L,R}^\dagger (2))^{n}\big )\Big ) 
\nonumber \\[1ex]\label{Theta} 
&=&\frac{\pi}{D}\Big (\mathbf{1}
+\frac{i}{2S}\sum_{n=1}^{S-1}\cot (\frac{\pi}{S}n )
\big (\hat T_{L,R}(2n)-\hat T_{L,R}^\dagger (2n)\big )\Big )  
\;\;, \end{eqnarray}
where either index $L$ or $R$ applies. These results hold for states with $\hat T_{L,R}(2)|\psi\rangle\neq |\psi\rangle$, {\it cf.} Eqs.\,(\ref{HamiltonianCog})-(\ref{HamiltonianFin}), 
otherwise $\hat\Theta_{L,R}|\psi_0\rangle=0$.  
 
We also may represent this slowing-down, produced by multiples of $k_0$ updates followed by instantaneous translations of $2l_0$ sites to the left (right) of the right-movers (left-movers), by an effective dimensionless Hamiltonian 
$\hat H_{eff}'$: 
\begin{equation}\label{Hefftransl} 
\hat H_{eff}':=k_0\hat HT +l_0(\hat\Theta_L+\hat\Theta_R)D  
\;\;. \end{equation}
Since all three operators on the right-hand side commute, the $l_0$ translations could be arbitrarily distributed among the $k_0$ updates in the  corresponding unitary update operator $\hat U_{eff}'$. 

While the pictorial representation of the action of the 
Hamiltonians on the spin chain is straightforward, their  structure seems quite unfamiliar. We emphasize that they 
generate the deterministic evolution of the Ising spin chain 
that we introduced as an ontological `toy model'.  
Underlying spin exchange interactions here lead to hopping 
(of values of `classical' spins) from sites to either left or right next-to-nearest-neighbour sites. 


\subsection{Equations of motion} 

We modify the notation to rewrite a generic chain state 
$|\psi\rangle$ from Eq.\,(\ref{chainstate}):  
\begin{eqnarray}\label{psinew} 
|\psi\rangle 
&\equiv &|\{ s(2j-1)\},\{ s(2j)\}\rangle \;\;,
\\[1ex]\label{psinew1} 	
&\equiv &|\{ s^L(j)\},\{ s^R(j)\}\rangle 
\;\;, \end{eqnarray} 
for $j=1,\dots,S$\,; 
{\it i.e.}, the left- and right-moving spin variables are first grouped separately, with $s(j):=s_j$, and then the 
even and odd numbered sites on the chain are renumbered.  

Considering, for example, the slowed-down evolution of 
{\it \underline{Case A}} generated by 
$(\hat U^\dagger )^{l_0}\hat U^{k_0}$ in 
$k_0+l_0$ steps, the update of a 
state $|\psi_n\rangle$ at time $t=nT$ is described {\it exactly} by the following pair of equations: 
\begin{eqnarray}\label{Lmover} 
s^L_{n+k_0+l_0}(j)-s^L_n(j)&=&
s^L_n(j+[k_0-l_0])-s^L_n(j) \;\;, 
\\[1ex]\label{Rmover} 
s^R_{n+k_0+l_0}(j)-s^R_n(j)&=&
s^R_n(j-[k_0-l_0])-s^R_n(j) 
\;\;, \end{eqnarray}
for $j=1,\dots,S$\,; here we suitably subtracted identical terms on both sides, respectively. These equations show that for $k_0+l_0$ steps of length $T$ 
forward in time only $k_0-l_0$ spatial steps of length $D$  are performed, {\it i.e.} the slowing-down compared to the case with $l_0=0$. 

Now, let the discrete time coordinate $n$ and space coordinate $j$ vary as integer multiples of $k_0+l_0$ and $k_0-l_0$, respectively, {\it i.e.} 
$n\equiv \tilde n(k_0+l_0)$ and $j\equiv \tilde j(k_0-l_0)$. Then, implicitly assuming redefined  
dimensional time and spatial scales  
$\tilde T:=(k_0+l_0)T$ and $\tilde D:=(k_0-l_0)D$, 
respectively, the equations of motion (\ref{Lmover}) and (\ref{Rmover}) can be rewritten more simply: 
\begin{eqnarray}\label{Lmover1} 
s^L_{\tilde n+1}(\tilde j)-s^L_{\tilde n}(\tilde j)&=&
s^L_{\tilde n}(\tilde j+1)-s^L_{\tilde n}(\tilde j) \;\;, 
\\[1ex]\label{Rmover1} 
s^R_{\tilde n+1}(\tilde j)-s^R_{\tilde n}(\tilde j)&=&
s^R_{\tilde n}(\tilde j-1)-s^R_{\tilde n}(\tilde j) 
\;\;, \end{eqnarray}
which are recognized as discretized first order partial differential equations in 1+1 dimensions. 
We note that left- and right-movers do not interact and up and down spins are never flipped, so far. 

In the continuum limit, introducing the linear combinations $S^\pm :=S^L\pm S^R$ and 
$\bar\sigma^0:=\mathbf{1}_2$, $\bar\sigma^1:=-\sigma_x$ with unit matrix $\mathbf{1}_2$ and Pauli matrix $\sigma_x$, the Eqs.\,(\ref{Lmover1})-(\ref{Rmover1}) 
can be combined in the form of the left-handed Weyl equation: 
\begin{equation}\label{LWeyl} 
\bar{\sigma}^\mu\partial_\mu\Psi_L
=\left (\mathbf{1}_2\partial_t-\sigma_x\partial_x\right )\Psi_L=0 
\;\;, \end{equation}
for a two-component `spinor' $(\Psi_L)^t:=(S^+,S^-)$. -- Correspondingly, the right-handed equation, with  
$\sigma^0:=\mathbf{1}_2$, $\sigma^1:=+\sigma_x$, and $(\Psi_R)^t:=(S^+,S^-)$,  
\begin{equation}\label{RWeyl} 
\sigma^\mu\partial_\mu\Psi_R
=\left (\mathbf{1}_2\partial_t+\sigma_x\partial_x\right )\Psi_R=0 
\;\;, \end{equation}
can be obtained by 
replacing $\partial_x\rightarrow -\partial_x$ in the derivation following from  Eqs.\,(\ref{Lmover1})-(\ref{Rmover1}); this amounts to 
an additional overall minus sign on the right-hand sides of these equations, {\it i.e.} exchange of the roles of left- and right-movers. The latter corresponds to numbering the sites of the spin chain from right to left 
instead of from left to right, as we did.  

Note that no mass terms appear in these equations. Nevertheless, if we keep track explicitly of the factors $k_0\pm l_0$ from Eqs.\,(\ref{Lmover})-(\ref{Rmover}), which were relevant for slowed-down evolution as compared to the one with signal velocity $c=1$ ({\it i.e.} $k_0=1$, $l_0=0$ in \underline{\it Case A} in subsection\,3.3.), we find that it is replaced by 
$c'=(k_0-l_0)/(k_0+l_0)$ (for $k_0>1$, $l_0>0$ and equal discreteness scales, $T=D$) here. 

Furthermore, if we apply {\it Sampling Theory},   
our {\it finite differences equations} are mapped one-to-one on continuous space-time 
partial differential equations for bandwidth-limited  fields $s^L(x,t)$ and $s^R(x,t)$, similarly  
as in Refs.\,\cite{PRA2014,Shannon,Jerri,Kempf}. 
In this way, one learns that Eqs.\,(\ref{LWeyl})-(\ref{RWeyl}) represent  continuum equations for functions with an ultraviolet cut-off corresponding to nonzero scales 
$T$ or $D$. More generally, this incorporates corrections in terms of higher-order derivatives (from expanding $\exp (T\partial_t)$ or $\exp (D\partial_x)$). Furthermore, on  discrete $x,t$ lattice points the functions can only assume the Ising spin values $s_k=\pm 1$, {\it cf.} Eq.\,(\ref{Isingspins}), while they can be real-valued in between.\,\footnote{This limitation can be overcome, as discussed in Section\,4. for the Dirac equation.} 


\subsection{Discussion -- the effective quantumness} 

Earlier we argued -- {\it e.g.} in Ref.\,\cite{Elze20} --  
that classical Ising spin chain models lend themselves 
to become quantum mechanical `by mistake', {\it i.e.} by 
slight imprecision of numerical constants appearing 
in their respective Hamiltonian. 

To remind us of this, we recall the relation between exchange operations (permutations) and Pauli matrices, 
$\hat P_{ij}=(\underline{\hat\sigma}_i\cdot\underline{\hat\sigma}_j+\mathbf{1})/2$. Let the Ising spins, on which  
update operator $\hat U$ and Hamiltonian $\hat H$ act, 
{\it cf.} Eqs.\,(\ref{Uchain}) and (\ref{HamiltonianCog})-(\ref{HamiltonianFin}), be embedded into a Hilbert space of 
corresponding qubits. Then $\hat H$ expressed in terms of Pauli matrices appears like a {\it quantum mechanical operator}.  

This implies that $\hat U$ or $\hat H$ 
describe the 
deterministic {\it evolution of ontological states} 
only in their precisely determined form:  
{\it no superpositions} of $\cal OS$ are produced, in agreement with the discussion in Section\,2. 

However, small inaccuracies in the form of these operators 
or their numerical `coupling' constants will produce unphysical superpositions of ontological states instead.    

This can be illustrated simply by the example of an exponential of an exchange operator, which itself does 
not produce superpositions:  
$$ 
i\exp (-i\frac{\pi}{2}(1+\epsilon )\hat P_{ij}) 
=\hat P_{ij}-i\frac{\pi}{2}\epsilon\cdot\mathbf{1}
+\mbox{O}(\epsilon^2)\;\;,\;\;
0<\epsilon\ll 1\;\;.$$ 
The resulting sum of terms unavoidably creates superposition states in a Hilbert space for qubits rather than bits or ontological Ising spins. In the same way, any approximation or perturbation of an Ising spin chain Hamiltonian is likely and surprisingly bound to produce quantum mechanical effects.   

Therefore, studying available Hamiltonians and laws of evolution -- which present some kind of approximations to possibly underlying ontological ones -- must suggest the most {\it effective interpretation} that the objects of interest behave in experiments as if of genuinely {\it quantum mechanical nature}, which is witnessed by the ubiquitous appearance of superposition 
states.\,\footnote{No doubt,   
available theories following quantum mechanics textbooks are very efficient 
and the most accurate ones in physics to this day, even though based on not 
so well understood foundations.}  


\section{A deterministic automaton, mass, and the Dirac equation} 

We have seen in subsection\,3.4. that a 
seemingly natural modification of the kinematics of free left- and right-movers does not yield a mass term in the resulting equations. It only has changed the signal velocity in the Ising spin chain model. Therefore, we address this 
problem here in another way, namely by ``reverse engineering", starting from the Dirac equation in 1+1 dimensions. 

We make use of a real representation 
of the Dirac equation studied in Ref.\,\cite{KauffmanNoyes}: 
\begin{equation}\label{KauffmanNoyesI}
\partial_t\Psi =
\left (\sigma_z\partial_x-i\mu\sigma_y\right )\Psi 
\,\,, \end{equation}
with a two-component real spinor field 
$\Psi^t:=(\Psi_1,\Psi_2)$\,; $\sigma_{y,z}$ are the usual Pauli matrices and $\mu$ is the mass parameter to be recovered in the spin chain model. 

More explicitly, the Eq.\,(\ref{KauffmanNoyesI}) gives 
two coupled partial differential equations:  
\begin{eqnarray}\label{Psi1} 
\partial_t\Psi_1&=&\partial_x\Psi_1-\mu\Psi_2
\;\;, \\ [1ex] \label{Psi2} 
\partial_t\Psi_2&=&-\partial_x\Psi_2+\mu\Psi_1 
\;\;. \end{eqnarray} 
It is tempting to discretize them, with the aim to go back to Eqs.\,(\ref{Lmover})-(\ref{Rmover}), 
however, in presence of mass terms. 

For this purpose we replace $\Psi_1(x,t)\rightarrow s_n^L(j)$ and 
$\Psi_2(x,t)\rightarrow s_n^R(j)$\,, {\it i.e.} in terms 
of left- and right-movers, respectively. Thus, 
instead of Eqs.\,(\ref{Lmover})-(\ref{Rmover}), for   
$k_0=1$ and $l_0=0$, we obtain: 
\begin{eqnarray}\label{Lmoverm} 
s^L_{n+1}(2j-1)&=&
s^L_n(2j+1)-\mu s_n^R(2j) \;\;, 
\\[1ex]\label{Rmoverm} 
s^R_{n+1}(2j)&=&
s^R_n(2j-2)+\mu s_n^L(2j-1)  
\;\;, \end{eqnarray}
for $j=1,\dots,S$\,, and cancelling identical terms on both sides of the 
equations that arise from discretization of the derivatives. Herein,  
the notation is as earlier, before  
modifying it in  Eqs.\,(\ref{psinew})-(\ref{psinew1}): the left-movers (right-movers) occupy the odd (even) numbered sites of the chain ({\it cf.} Figure\,1\,a)). 

We note that the mass terms couple the left- and right-movers. This changes profoundly the character of the chain model. Due to their additive contributions, 
the variables generally can no longer be restricted to represent 
two-state Ising spins. Choosing $\mu =1$ (more generally, integer valued) the variables can, however, be maintained to be integers. 

In order to allow the spinor field to become a real number and its absolute value possibly large in the continuum limit, it is necessary to replace two-state Ising spins by generalized $(2M+1)$-state objects, with $M$ a sufficiently large but finite natural number. In the end discrete quantities can then be mapped to continuous ones by invoking {\it Sampling Theory}, as we  
recalled at the end of subsection\,3.4.  


\subsection{Some arithmetic with permutations of discrete states} 
 
In order to accommodate the additive contributions of the mass terms 
in Eqs.\,(\ref{Lmoverm})-(\ref{Rmoverm}), we invoke once more a periodic compactification of the state space ({\it cf.} the Cogwheel Model of subsection\,3.1.). Thus, the generalized variables can assume the values 
$-M\leq s\leq +M$ and addition/subtraction due to the mass terms has to respect  
identification at the boundary $M+1\equiv -M$ or $-M-1\equiv +M$. 
 
Consider the Eq.\,(\ref{Rmoverm}), for example. Our aim is to show that the addition of the two terms on the right-hand side can be implemented in terms of permutations. This will demonstrate that these automaton equations 
qualify to describe deterministic evolution of ontological states by 
permutations, as required by the {\it Cellular Automaton Interpretation} introduced in Sections\,1. and 2. 

For illustration, the simplest possible case with $M=1$ has     
classical three-state spin variables, $s_n(j)$ at chain site $j$ and instant $n$, which always assume exactly one of the values $s^{m=1,2,3}\equiv -1,0,+1$\,. All possible results of addition and subtraction 
($\mu =1$), $S^R+S^L$ in Eq.\,(\ref{Rmoverm}) and $S^L-S^R$ in Eq.\,(\ref{Lmoverm}), respectively, 
are then collected by symbolic matrices $\hat{\cal S}_+$ and $\hat{\cal S}_-$: 
\begin{center} 
\begin{tabular}{r | c c c c} \label{table} 
      & $s^1$ & $s^2$ & $s^3$ & $S^R$ \\ \hline 
$s^1$ & $s^3$ & $s^1$ & $s^2$ &       \\ 
$s^2$ & $s^1$ & $s^2$ & $s^3$ &       \\ 
$s^3$ & $s^2$ & $s^3$ & $s^1$ &       \\ 
$S^L$ &       &       &       &       \\ 
\end{tabular} 
$=:\hat{\cal S}_+\;\;$,\,\,\,\,\, 
\begin{tabular}{r | c c c c} \label{table} 
      & $s^1$ & $s^2$ & $s^3$ & $S^L$ \\ \hline 
$s^1$ & $s^2$ & $s^3$ & $s^1$ &       \\ 
$s^2$ & $s^1$ & $s^2$ & $s^3$ &       \\ 
$s^3$ & $s^3$ & $s^1$ & $s^2$ &       \\ 
$S^R$ &       &       &       &       \\ 
\end{tabular} 
$=:\hat{\cal S}_-\;\;$.  
\end{center}    
The matrix $\hat{\cal S}_+$ ($\hat{\cal S}_-$) is symmetric (antisymmetric) and shows a nice permutation structure, caused by addition (subtraction) obeying the  identification of boundary values, $M+1\equiv -M$ or $-M-1\equiv +M$, for $M=1$ in this example. This can be read as follows. The value of the mass terms (left vertical) 
conditions which permutation composed of transpositions ({\it cf.}  Eqs.\,(\ref{spinexchange})-(\ref{exchangeprop})), applied to the set of values of $S^{R,L}$ (top horizontal), 
gives the corresponding row of the matrix. {\it E.g.}, 
$S^L=s^3$ can be associated with the permutation $\Pi_+^3$,   
$$s^3\;\rightarrow\;\Pi_+^3\{ s^1\; s^2\; s^3\}:=\hat P^{13}\hat P^{23}\{ s^1\; s^2\; s^3\}
=\{ s^2\; s^3\; s^1\}
\;\;, $$ 
which gives the third row of $\hat{\cal S}_+$, {\it etc.} 

One may identify the symbols for the three 
spin states with standard basis vectors: 
$s^1\equiv (1,0,0)^t=:\vec s^{\;1}$, $s^2\equiv (0,1,0)^t=:\vec s^{\;2}$, 
$s^3\equiv (0,0,1)^t=:\vec s^{\;3}$. The result of an addition, such as $S^R+S^L=-1+1$, can then always be expressed in terms of these 
vectors and the matrix $\hat{\cal S}_+$: 
$S^R+S^L=s^1+s^3=\vec s^{\;1}\cdot\hat{\cal S}_+\cdot\vec s^{\;3}=s^2=0$. 
Similarly, subtraction is expressed with the help of $\hat{\cal S}_-$ 
and these basis vectors. 

This may seem unnecessarily tedious. However, it illustrates that addition and subtraction on a finite set of adjacent numbers with `periodic boundary condition', such as $-1,0,+1$ with $2\equiv -1$, $-2\equiv 1$, can be obtained from permutations of these numbers, which represent discrete spin states. 

These considerations are obviously not limited to the example of a small number of spin states, $2M+1$. 

To summarize, a discrete deterministic 
automaton representation for the Dirac equation in 
1+1 dimensions (with integer mass) is given by Eqs.\,(\ref{Lmoverm})-(\ref{Rmoverm}), which describe 
a chain of ($2M+1$)-state Ising spins with periodic 
boundary condition. 
The states beyond the lowest ($-M$) and highest ($+M$) states have to be periodically identified, as for a finite set of states compactified 
on a circle. The discrete equations, the mass terms 
in particular, mix the left- and right-movers that 
we studied in subsection\,3.3.-3.4. and change the 
spin state at each site simultaneously. -- 
The overall picture to keep in mind is that of a  
{\it ``Necklace of Necklaces''}.  

It will be interesting to compose the related  
unitary update operator $\hat U$ explicitly from 
permutations (kinematics of 
left- and right-movers), similarly as before in subsection\,3.2., and the permutations acting on the symbols representing discrete states at a given site, as discussed above, which change the spin 
states.\,\footnote{We remark that  
pairwise spin exchange operators can be calculated 
as polynomials in the product of spin operators,  
$\vec S_i\cdot\vec S_j$, also for multi-state spins, 
{\it i.e.} large representations with $M\gg 1$.\,\cite{HABrown}} 
This will be presented elsewhere. 


\section{Conclusions} 

A classical discrete deterministic Ising spin model with nearest-neighbour interactions in the form of spin exchange permutations serves us to illustrate in detail the cellular automaton ontology proposed by G.\,'t\,Hooft to underlie 
quantum theory.\,\cite{tHooft2014} We argue in particular that this model of classical two-state 
entities is bound to become quantum mechanical, by creating superposition states of qubits, if it is deformed in the slightest way. All parameters of such model 
seem to be exactly fixed by its permutation dynamics. 

We study whether and how the finite signal velocity in such a spin chain can be modified. However, we find that this kind of obvious modifications does not relate 
to mass terms. In fact, in the continuum limit this model represents the Weyl equation describing massless particles. 

Nevertheless, we show that mass can be incorporated, in a way motivated by the 
Dirac equation in 1+1 dimensions. Interestingly our cellular automaton in this 
case, based on permutations of real integer variables, seems simpler than many of the {\it Checkerboard Models}, following Feynman, 
which try to discretize the path contributions to the 
propagator of the Dirac equation, see Refs.\,\cite{Feynman,Iwo,Ord} and references therein. They always presuppose quantum mechanical effects, 
such as superposition of complex amplitudes. The generalization of our approach -- motivated instead by the search for ontological dynamics based on permutations -- remains to be seen. 

Finally, in Section\,5. we did not follow up on the argument of subsection\,3.5., concerning classical two-state Ising spins that become quantum mechanical `by mistake', which led quite   
naturally to a model for qubits. This required permutations and evolution of discrete ontological states to be related to quantum mechanical operators, the Pauli 
matrices. For the case of the Dirac equation this does not seem warranted. ``Effective  quantization'' as in 3.5., deforming its discrete automaton somehow, does not reproduce in any obvious way   
the quantum field theory of free fermions which incorporates the Pauli principle. We leave this to be understood. 


\section*{Acknowledgments} 

It is a pleasure to thank Ken Konishi for 
many discussions of foundations of quantum 
theory and Marco Genovese and Paolo Oliveiro for kind invitation to 
{\it Quantum 2023 (Torino, September 10-15, 2023)}, 
where part of this work was presented. An anonymous 
supportive referee is thanked for suggesting additional references.  




\end{document}